\documentstyle[floats,aps,prb]{revtex}

\input epsf

\newcommand{\ket}[1]{\mid #1 \rangle}
\newcommand{\half}{{\textstyle \frac 1 2}}

\begin{document}
%\fontsize{12pt}{12pt}
\selectfont
\twocolumn[\hsize\textwidth\columnwidth\hsize\csname@twocolumnfalse\endcsname

\title{Inter-Landau-level skyrmions versus quasielectrons in the
  $\nu=2$ quantum Hall effect}
\author{D. Lillieh\"o\"ok}

\address{Department of Physics, Stockholm University, Box 6730,
  S-11385 Stockholm, Sweden }

\date{\today} \maketitle

\begin{abstract}
  {We consider the charged excitations in the quantum Hall effect
    with a large Zeeman splitting at filling factor $\nu=2$. When
    the Zeeman splitting is increased over a critical value the
    ground state undergoes a first order phase transition from a
    paramagnetic to a ferromagnetic phase. We have studied the
    possibility of forming charged spin-textured excitations,
    ``inter-Landau-level skyrmions,'' close to this transition,
    and find that these never are the lowest lying charged
    excitations but can under certain conditions provide an
    effective driving mechanism for the paramagnetic to
    ferromagnetic phase transition.  We show that the charged
    excitations inevitably present in the system can act as
    nucleation centers even at $T=0$ and hence set a specific
    limit for the maximal hysteresis attainable in this case.
    Calculations of how the finite width of the two-dimensional
    electron gas affects the $\nu=2$ phase transition and the
    polarization at higher filling factors are also presented.}
\end{abstract}

\vskip 2pc

] \narrowtext

\section{Introduction}
Ferromagnetic quantum Hall systems are known to have charged
excitations, skyrmions, that involve texturing of spin. So far,
these skyrmions have been considered only at odd integer filling
factors and fractional filling factors smaller than 1.
Skyrmions were predicted to be the lowest energy charged
excitations in, for example, GaAs at filling factor $\nu=1$.
\cite{sondhi1,fertig} This was later confirmed in several
experiments. \cite{barrett,schmeller,goldberg,eaves} Under some
conditions skyrmions are also predicted to form at higher odd
filling factors \cite{Cooper,sondhi2} and there are experiments
claiming to have seen this. \cite{nu3experiment}

Since skyrmions involve moving particles to unoccupied
spin-flipped states, they can be candidates for the lowest energy
excitations only when there is a small single-particle gap for
such spin flips. In most materials spin-orbit coupling decreases
the effective Land\'e factor $g$ of the electrons---in for
example GaAs $g$ is effectively reduced from 2 to $-0.44$.
At the same time the cyclotron gap $\hbar \omega_c$ is normally
increased due to the small effective mass of the conducting
electrons in these materials. This means that the Zeeman energy
$g \mu_B B$ will normally be small compared to the cyclotron
gap $\hbar \omega_c$. Hence small single-particle gaps will be
found at odd filling factors.

There are, however, materials where spin-orbit coupling is so
strong that the magnitude of the Zeeman energy is instead
strongly enhanced. In InSb the effective Land\'e factor $g$ is of
the order of $-50$. \cite{Merkt} The ratio of the Zeeman energy
to the cyclotron energy gap can also, as always, be increased by
tilting the sample relative to the magnetic field. This makes it
possible to reach a limit where the spin-split single-particle
energy levels from the lowest and next lowest orbital Landau
levels can come very close together or even cross (see
Fig.~\ref{LargeZeemanFig}). In this case there is a small
single-particle gap at {\em even} filling factors and one can
imagine the possibility of having ``inter-Landau-level''
skyrmions---these would be charged excitations that involve
spins flipped from one orbital Landau level to another.

In this paper we first review the paramagnetic to ferromagnetic
phase transition that occurs at filling factor $\nu = 2$ when
single-particle levels from different orbital Landau levels come
close. We present finite width calculations of how the Coulomb
interaction affects this transition and comment on recent
experiments at higher filling factors by Papadakis {\em et al.\ 
  }\cite{Shayegan} on the spin splitting in AlAs quantum wells.
We then investigate the charged excitations in the possible
$\nu=2$ quantum Hall states and find that inter-Landau-level
skyrmions are never the lowest energy charged excitations in this
system, but do provide an effective driving mechanism for the
phase transition and thereby limit the maximal hysteresis
attainable in this case.

\section{The Paramagnetic to Ferromagnetic Transition}
In the limit when the Coulomb energy is negligible compared to
the gaps in the single-particle spectrum, we can specify the
groundstate of a quantum Hall system at integer filling factors
by giving the order in which the single-particle energy levels
\begin{figure}[htbp]
  \centerline{\epsfbox{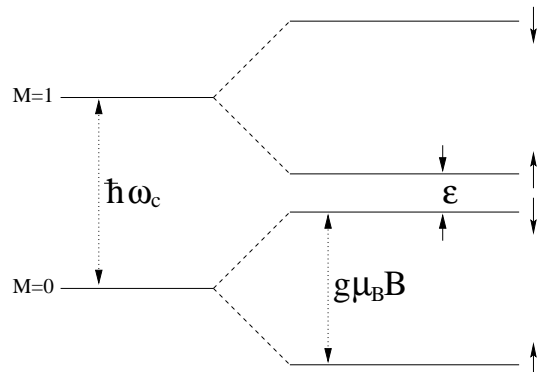}}
  \caption{The single-particle energy spectrum. Here the Zeeman gap
    $g \mu_B B$ and the cyclotron gap $\hbar \omega_c$ are
    comparable in magnitude, leaving a small gap $\epsilon$
    between the second and third levels.}
  \label{LargeZeemanFig}
\end{figure}
are filled. For finite but small Coulomb interaction, spin
polarized states will be favored and the order in which levels
are filled is determined by a competition between single-particle
energies and the Coulomb energy.

In the case when the cyclotron gap $\hbar \omega_c$ is larger
than both the Zeeman energy $g \mu_B B$ and the characteristic
Coulomb energy $e^2/ \epsilon \ell$, the $\nu=2$ ground state is
simply the lowest orbital Landau level with both spin states
filled. This state is paramagnetic with no net polarization. Now
if the Zeeman energy is increased the system will undergo a first
order phase transition to the ferromagnetic state with the same
spin filled in the two lowest Landau levels.  The Coulomb
interaction will cause this transition to occur {\em before} the
single-particle energies cross. \cite{Giuliani,MacDonald98}
Evaluating the Coulomb energy in the paramagnetic state
$E^{\mbox{\footnotesize PM}}_C$ and the ferromagnetic state
$E^{\mbox{\footnotesize FM}}_C$ yields
\begin{eqnarray}
  E^{\mbox{\footnotesize PM}}_C&=&2 E_{00} , \nonumber \\
  E^{\mbox{\footnotesize FM}}_C&=& E_{00} +E_{11} +2 E_{01} .
\end{eqnarray}
Here $E_{MN}$ is the total exchange energy contribution of the
filled level $M$ to level $N$,
\begin{eqnarray}
  E_{MN} = &-& \half \sum_{m,n}
  \int d^2r \int d^2r' V(|{\bf r}-{\bf r}'|) \times 
  \nonumber \\ &&
  \phi^*_{Mm}({\bf r}) \phi_{Nn}({\bf r}) 
  \phi^*_{Nn}({\bf r}') \phi_{Mm}({\bf r}') \, ,
\end{eqnarray}
where $\phi_{Mm}$ is the wave function with (angular) momentum
$m$ in Landau level $M$. In the ideal (zero thickness)
two-dimensional case $E_{MN}$ can be evaluated exactly,
\begin{equation}
  E_{MN} = -N_\phi \frac{1}{\sqrt{2}} \int_0^\infty dt e^{-t^2} 
  L_M(t^2) L_N(t^2) ,
\end{equation}
where $L_M$ are the Laguerre polynomials and $N_\phi$ is the
number of flux quanta in the system, i.e., the number of
particles in each level. The difference in Coulomb energy between
the two states, $E^{\mbox{\footnotesize PM}}_C -
E^{\mbox{\footnotesize FM}}_C,$ in this ideal two-dimensional
(2D) case is $\frac{3}{8}\sqrt{\pi/2} N_{\phi}$; this
means that the ferromagnetic state will have the lowest total
energy when
\begin{equation}
  \epsilon \equiv \hbar \omega_c - g \mu_B B \leq \epsilon_c=
  \frac{3}{8}\sqrt{\frac{\pi}{2}} \approx 0.470 ,
\end{equation}
where energies here and onward are given in units of
$e^2 / \epsilon \ell$.
This instability may seem large but is drastically reduced when
the finite width of the 2D electron gas is taken into account.
Using a Gaussian subband approximation \cite{Cooper} we calculate
$\epsilon_c$ numerically for different effective widths $w$.
Results are given in Table~\ref{ew} where $w$ is given in units
of the magnetic length.
\begin{table}[ht]
  \begin{tabular}[ht]{lllll}
       w & 0 & 0.5 &  1.0 & 2.0 \\
    $\epsilon_c$ & 0.4700 & 
    0.2010 & 0.1227 & 0.0673 
  \end{tabular}
  \caption{Finite width dependence of $\epsilon_c$.}
  \label{ew}
\end{table}
The finite width dependence of $\epsilon_c$ closely follows the
form
\begin{equation}
  \epsilon_c(w) = \frac{a}{w+b},
  \label{nu2gap}
\end{equation}
where $a \approx 0.170$ and $b \approx 0.359$.

It is worth noting that for small enough densities (and hence
small magnetic fields) the ferromagnetic $\nu=2$ state has the
lowest energy even for a vanishing Zeeman energy. This happens
because the Coulomb energy---proportional to $\sqrt B$---is
larger than the cyclotron energy---proportional to $B$---for
small enough $B$. Again the critical density for this
ferromagnetic ordering depends strongly on the width of the 2D
gas. We can easily predict it for some common materials, now
keeping the Zeeman energy at zero tilt angle at $\nu=2$; see
Table~\ref{fm2}.
\begin{table}[h]
  \begin{tabular}[h]{lllll}
    & & \hspace{1cm} w \\
    & 0 & 0.5 & 1.0 & 2.0 \\
    \hline
    InSb \rule{0cm}{0.40cm} & $4.8\times10^9$ & $8.8\times10^8$ 
    & $3.3\times 10^8$ & $9.9\times 10^7$ \\
    GaAs & $7.5 \times 10^{10}$ & $1.4 \times 10^{10}$ & $5.1 \times 10^9$
    & $1.5 \times 10^9$ \\
    AlAs & $2.4 \times 10^{13}$ & $4.4 \times 10^{12}$ & $1.6 \times 10^{12}$ 
    & $4.9 \times 10^{11}$
  \end{tabular}
  \label{fm2}
  \caption{Critical densities (in cm$^{-2}$) below which the $\nu=2$ ground 
    state will be polarized at zero tilt angle for different widths of the 
    2D electron gas.}
\end{table}

In GaAs and InSb these critical densities are too low to be
experimentally relevant. In AlAs we use an effective mass of
$m^*=0.46m_e$ corresponding to the 2D electron gas occupying
ellipsoidic constant-energy surfaces with one major axis in the
plane of the 2D interface and find a critical density that is
larger than that used in experiments. This is compatible with
experiments by Papadakis {\em et al.} \cite{Shayegan} who report
on measurements where they see a fully polarized $\nu=3$ state at
zero tilt angle in this system. We must note that, since the
Coulomb energy normally is larger than the cyclotron gap in AlAs,
the true ground state will contain a large mixture of different
Landau levels and the simple analysis here is not expected to be
precise. Nevertheless, the same method applied to the $\nu=3$
ground state with a physical width of 150 \AA \ (as used in the
setup of Papadakis {\em et. al.}) yields a critical density of
$3.8\times10^{11}$ cm$^{-2}$ below which the $\nu=3$ groundstate
should be fully polarized. This agrees qualitatively with the
range of densities used in this experiment
[(1.4--3.9)$\times10^{11}$~cm$^{-2}$].

\section{The charged excitations}
We have used a time dependent Hartree-Fock method to look for
spin-flip instabilities around a quasielectron or quasihole in
both the paramagnetic and ferromagnetic $\nu=2$ ground states.
This instability analysis is equivalent to taking the small spin
limit of an inter-Landau-level skyrmion and will tell us whether
or not the charged excitations involve extra flipped spins.
Before describing the calculation we want to note that Kohn's
theorem, \cite{Kohn} stating that electron-electron interactions
cannot affect the cyclotron resonance in a translationally
invariant system, does not imply anything for the excitations in
this case since Kohn's theorem does not apply when the excited
state has a different spin from
%than???
the ground state.

In the paramagnetic case we write the ground state as
\begin{equation}
  \ket{\psi_0}_{\mbox{\footnotesize PM}}=\prod_{m=0} 
  c^\dagger_{0,m,\downarrow}
  c^\dagger_{0,m,\uparrow} \ket{0} ,
\end{equation}
where $c^\dagger_{M,m,\sigma}$ creates an electron in Landau
level $M$ with angular momentum $m$ and spin $\sigma$. We use the
symmetric gauge where $m$ takes integer values from $-M$ and
upward.  Next we create a charged excitation (a quasihole) at
the origin by removing one electron from the upper level. To find
out if further spin flips are favored we allow an inter Landau
level spin wave around the hole,
\begin{equation}
  \Psi^\dagger(q) \ c_{0,0,\downarrow} 
  \ket{\psi_0}_{\mbox{\footnotesize PM}} ,
\end{equation}
where $\Psi^\dagger(q)$ is a spin wave operator that flips a spin
from the $M=0$ spin-down level to the $M=1$ spin-up level with
change $q$ in momentum,
\begin{equation}
  \Psi^\dagger\left(q\right)= \sum_{k=1} \alpha_k 
  c^\dagger_{1,k+q,\uparrow} c_{0,k,\downarrow} .
\end{equation}
We determine the $\alpha_k$'s and the corresponding energies by
numerically diagonalizing the Coulomb Hamiltonian in this
subspace for each $q$.  Note that $q$ can take values from $-2$
and upward here.  For $q=-2$ we find a negative eigenvalue of
$-0.269$ in the ideal 2D case. This means that the charged
excitations will involve extra spin flips and form a charged
spin-texture excitation---a skyrmion---if the single-particle gap
to the next level is smaller than this instability, i.e.,
$\epsilon \leq 0.269 e^2/\epsilon\ell$.  However, this value is
smaller than the instability to the ferromagnetic state described
in the previous section ($\epsilon_c=0.470$); thus the system
will in this case undergo the transition to the fully polarized
state {\em before} skyrmions become the preferred quasiparticles.

Since this is a question of energetics it is not guaranteed that
the same conclusion holds for finite widths of the 2D electron
gas. Indeed, at filling factor $\nu=3$ skyrmions are predicted to
form for finite widths but not in the ideal 2D case.
\cite{Cooper,sondhi2} In the present case, however, the two
instabilities both decrease at roughly the same rate and the
polarized electron/hole quasiparticle is favored up to a width of
at least five magnetic lengths.

In the ferromagnetic state,
\begin{equation}
 \ket{\psi_0}_{\mbox{\footnotesize FM}}=\prod_{m=0, n=-1}
c^\dagger_{1,n,\uparrow} c^\dagger_{0,m,\uparrow} \ket{0} ,
\end{equation}
the system can {\em gain} single-particle energy by flipping one
spin from the filled $M=1$ spin-up level to the empty $M=0$
spin-down level below it.  By first diagonalizing the Coulomb
Hamiltonian of this spin wave with no hole present, we confirm
that any such spin flip always costs more than the maximum
possible single-particle gain---the lowest energy spinwave costs
$0.51e^2/\epsilon\ell$. Hence the ground state is stable and the
phase transition is first order as promised.  We then redo the
diagonalization with a quasihole present and find that the spin
wave is unaffected by the localized excitation, so in this case
the spin wave again costs more Coulomb energy than the
potential single-particle gain. Hence there is no situation where
a skyrmion can have lower energy than an electron/hole
quasiparticle in this case either.

Since the transition between the paramagnetic and ferromagnetic
phase is first order, it should be possible to ``supercool'' the
system and stay in the metastable state. One could imagine, for
example, starting in the stable paramagnetic phase and then
slowly increase the Zeeman splitting while keeping the filling
factor fixed close to $\nu=2$.  This could be realized by
rotating the sample {\em in situ} while increasing the total
magnetic field.  When the single-particle gap $\epsilon$ falls
below $\epsilon_c$ ($=0.47$ in the 2D case) the paramagnetic
state ceases to be the true ground state.  However, there may
still be no effective way for the system to overcome the barrier
separating the two phases, in which case it will stay in the
metastable paramagnetic phase for a very long time. In a region
below $\epsilon_c$ the barrier is large since a single spin flip
still costs a large energy (even though we know that flipping
{\em all} spins would take us to the true ground state).
Naively, one could hope to form skyrmions by supercooling the
paramagnetic state beyond the point where the skyrmion
instability sets in (i.e., $\epsilon < 0.269$). But since forming
a skyrmion does involve flipping spins it seems likely that
instead of forming skyrmions in the metastable paramagnetic state
this spin-flip instability rather provides an effective way for
the system to undergo the phase transition and actually drives it
to the true (ferromagnetic) ground state. Note that this
instability will be present even at $T=0$ since there will always
be a finite density of charged excitations in the system (except
at the exact center of the $\nu=2$ plateaux). Once the
instability sets in, these charged excitations will act as
nucleation centers for the other phase and thus this instability
sets an upper limit for how large a hysteresis one can obtain
when approaching $T=0$.

We can also note that the hysteresis is asymmetric in the sense
that the paramagnetic state can survive as metastable further
into the ferromagnetic region than the ferromagnetic state can
into the paramagnetic region (the hysteresis is limited to $0.269
< \epsilon < 0.51$, with $\epsilon_c=0.47$ rather close to the
upper of these limits).

\section{Conclusion}
We have investigated the nature of the charged excitations in a
quantum Hall system with large Zeeman energy at filling factor
$\nu=2$, and find that there is no instability to flip extra
spins around the polarized quasiparticles either in the
paramagnetic or in the ferromagnetic ground state. Hence no
inter-Landau-level skyrmions can be the lowest energy charged
excitations in this system. The skyrmion instability does,
however, limit the possible hysteresis attainable in the
paramagnetic to ferromagnetic phase transition. We also presented
finite thickness calculations for the first order phase
transition between the paramagnetic and the ferromagnetic ground
state and predict some critical densities below which the
ferromagnetic state has lowest energy even for zero tilt angle.

\section{Acknowledgments}
The author wishes to thank Anders Karlhede, Kennet Lejnell, and
Hans Hansson for numerous useful discussions. The author also
thanks K. Mullen, N. Cooper, H. Fertig, G. Khodaparast, and S. Das
Sarma.

\end{document}